%% file: main.tex
\documentclass[sigconf]{acmart}

\AtBeginDocument{%
  \providecommand\BibTeX{{%
    \normalfont B\kern-0.5em{\scshape i\kern-0.25em b}\kern-0.8em\TeX}}}

\setcopyright{acmcopyright}
\copyrightyear{2021}
\acmYear{2021}
\acmDOI{10.1145/1122445.1122456}

\acmConference[Yokohama '21]{Woodstock '18: ACM CHI Conference on Human Factors in Computing Systems}{May 08--13, 2021}{Yokohama, Kanagawa}
\acmBooktitle{Yokohama '21: ACM CHI Conference on Human Factors in Computing Systems, May 08--13, 2021, Yokohama, Kanagawa}
\acmPrice{15.00}
\acmISBN{978-1-4503-XXXX-X/18/06}
\newcommand{\etal}{\textit{et al.}}

\begin{document}

\title{Exploring a Makeup Support System for Transgender Passing based on Automatic Gender Recognition}

\author{Toby Chong}
\authornote{Both authors contributed equally to this research.}

\affiliation{%
  \institution{The University of Tokyo}
  \city{Tokyo}
  \country{Japan}
}

\author{Nolwenn Maudet}
\authornotemark[1]
\affiliation{%
  \institution{University of Strasbourg}
  \city{Strasbourg}
  \country{France}
}

\author{Katsuki Harima}
\affiliation{%
 \institution{Harima mental clinic}
 \city{Tokyo}
 \country{Japan}
}

\author{Takeo Igarashi}
\affiliation{\institution{The University of Tokyo}
 \city{Tokyo}
 \country{Japan}
}

\renewcommand{\shortauthors}{Toby and Nolwenn, et al.}

\begin{abstract}
\input{0-abstract.tex}
\end{abstract}




\maketitle

\input{1-intro.tex}
\input{2-related_work.tex}

\input{3-design_process.tex}
\input{4-implementation.tex}
\input{5-user_study.tex}

\input{6-findings_makeup.tex}
\input{7-findings_system.tex}
\input{8-discussion.tex}
\input{9-conclusion.tex}
\input{10-acknowledgement.tex}
\bibliographystyle{ACM-Reference-Format}
\bibliography{sample-base}


\end{document}

%% file: 0-abstract.tex
How to handle gender with machine learning is a controversial topic. A growing critical body of research brought attention to the numerous issues transgender communities face with the adoption of current automatic gender recognition (AGR) systems. In contrast, we explore how such technologies could potentially be appropriated to support transgender practices and needs, especially in non-Western contexts like Japan.
We designed a virtual makeup probe to assist transgender individuals with passing, that is to be perceived as the gender they identify as. To understand how such an application might support expressing transgender individuals gender identity or not, we interviewed 15 individuals in Tokyo and found that in the right context and under strict conditions, AGR based systems could assist transgender passing.

Project Page: \url{https://sites.google.com/view/flyingcolor}

%% file: 1-intro.tex
\section{Introduction}
\label{sec:intro}

Among the different types of machine learning techniques, automatic gender recognition (AGR) systems, which probabilistically identify the gender of a person in a media (usually an image or a video clip) has become a standard part of image tagging services used by major tech companies such as Microsoft~\footnote{https://azure.microsoft.com/en-us/services/cognitive-services/face/} and SenseTime~\footnote{https://www.sensetime.com/en/Technology/face.html}. Anyone can upload an image to the service and get all faces in the image `gendered', that is, assigned a label of male or female, potentially alongside a confidence score, within milliseconds. These systems, often developed using data labeled as binary men and women exclusively, reportedly perform significantly worse on some populations, especially transgender individuals~\cite{scheuerman2019computers}. 
Unlike facial recognition (technology that matches a face with an identity from a database), which is facing ban in several states in the United States~\footnote{https://www.nytimes.com/2019/05/14/us/facial-recognition-ban-san-francisco.html}, image tagging services, which include AGR, are not yet being regulated under any jurisdiction, to the best of our knowledge. Nonetheless, when interviewed, transgender individuals in the United States expressed overwhelmingly negative attitudes toward AGR and its potential harmful applications~\cite{buolamwini2018gender}, prompting calls from researchers for severely restricting its use in order to protect transgender populations~\cite{keyes2018misgendering, chiagr2018}.

In contrast, work focusing on applying machine learning to mitigate gender bias or to support transgender in their daily life and cater to their specific needs has only started to appear recently ~\cite{voice_training}. 

\subsection{Motivation and Positionality}
Our long-term goal is to design techniques for mitigating gender bias and supporting the expression of gender identity. More specifically, we want to explore whether and how AGR could potentially be appropriated to support transgender individuals' specific practices and needs. This project was initially motivated by a 3-year long ethnography that revealed that many transgender women in Japan expressed difficulties online for learning how to apply makeup because of the lack of information available compared with the wealth of resources that target cis-woman. On the basis of this initial research as well as their HCI/CG/ML background, the first author proposed the initial idea of a virtual makeup system that supports transgender people's need. Building on this initial idea, we formed an interdisciplinary team of researchers of diverse gender identities and different cultural backgrounds (Japan, Hong Kong and France) to develop this project. In terms of positionality~\cite{doi:10.1080/23265507.2017.1300068} we also need to note that our experience with gender studies as well as transgender experiences is heavily influenced by our physical location: Japan.

Our second motivation for this work is to better document the perceptions and experiences with technologies of transgender populations outside of the United States and the Western world to foster a critical debate surrounding AGR. Most of the HCI studies focusing on transgender currently stem from the United States or other Western countries (e.g., ~\cite{SafeSpaces,ahmed2018trans, Haimson2020}). Although knowledge and insights gained from those studies might partially be applicable to transgender individuals from other cultural backgrounds, we argue, alongside many HCI researchers, that as a community we need to diversify the populations we are working with to let different voices be heard and acknowledged~\cite{kamppuri2006expanding, 10.1145/3025453.3025766}. 

\subsection{Studying transgender people in Japan}
This work is therefore situated in the specific context of transgender people in Japan. In this context, trans preoccupations do not always align with the ones identified by HCI researchers focusing on Western transgender experiences and needs (see \cite{yuen2018mediated, mackie2008girl, Michelle2020danso}). For example, in her account of female-to-male (FtM) transgender experiences in Tokyo, Yuen~\cite{yuen2020unqueer} explains that ``while some feminist and queer theorists have emphasized the transgressive potential of transgender — that by nature of their gender fluidity trans people can function as vehicles to challenge prevailing gender norms and overcome the limitations of the binaristic gender/sexual system — subverting or challenging the gender status-quo is not something many FtM trans people in Japan are concerned with''. 

In fact, in the Japanese context, researchers have shown the importance of passing for the many transgender who do not want to disclose their trans identity~\cite{yuen2018mediated}. For transgender people, passing is being perceived as the gender they wish to present as. For example, a transgender woman passes when other people who interact with her assume that she is a cis-female. In a Japanese context, Nishino~\cite{recontruct}, who interviewed $16$ Japanese transgender women pointed out that their gender dysphoria meant that they had trouble thinking that they looked like women. In a US context, Anderson~\cite{anderson2019your} found that among their participants who lived in Nebraska, motivations for passing include identity expression, affirmation, fear of misgendering and discrimination. 

It is critical to point out that while passing is often considered a critical part of one’s gender identity and a goal for many transgender individuals (e.g.,\ 92\% of women of color in the United States wish to pass or live in stealth~\cite{sevelius2013gender}), it is not a requirement for transgender individuals.
However, it is equally important to support trans women and men who wish to pass, regardless of their motivation, especially as it has been shown that a higher level of passing negatively correlates to the level of discrimination faced by the individuals, such as when receiving health care or looking for shelter when homeless~\cite{begun2016conforming, bockting2013stigma, kattari2016differences}. 

Today, trans people employ different techniques to pass and some of them are commonly associated with cisgender people in our society, such as makeup~\cite{anderson2019your}. 
This is especially true in the Japanese context where makeup is commonly expected from (trans) women~\cite{mackie2008girl}. 
In their paper discussing trans-technologies, Haimson~\etal~\cite{Haimson2020} explored technologies specifically designed by and for the trans community. Among the different types of applications envisioned, one category of ideas focused on technologies for changing appearance and technologies for gender affirmation. Although none of the ideas focused on makeup specifically, developing tools that support transgender individuals who wish to use makeup as a way to affirm their gender fits within this objective.

We therefore chose to create a makeup recommendation system specifically targeted at the Asian transgender community to help them explore how they can pass in an intimate and private setting.
In this paper, our contribution is threefold. We conducted a study with 15 transgender participants living in the greater Tokyo area to document Japanese transgender people's experience with passing and makeup in Japan. We propose an experimental system that leverages AGR to assist makeup exploration for passing dedicated to transgender people in Japan. We explored how participants perceive and use our system as well as understand whether and how we could potentially develop positive uses of AGR for the transgender community.

%% file: 2-related_work.tex
\section{Related Work}
\label{sec:related}
In this section, we start by discussing gender issues within machine learning in general. 
We then provide a brief overview of existing AGR technologies, including performance and applications, as well as critical studies that focused on how transgender people are currently being impacted by them. We also review existing research on virtual makeup applications, that is makeup applied onto picture or video of a face as well as recommendation systems that compute personalized makeup recommendation on the basis of a person's face.

\subsection{Gender in machine learning}
The relationship between gender and machine learning is a controversial topic. In \textit{Gender Shades}~\cite{buolamwini2018gender} for example, the authors revealed that commercial face recognition software show a large performance gap between light-skinned man and darker-skinned woman. In turn, a growing body of research has been discussing the wide range of issues associated with machine learning techniques and potential ways to mitigate them, including using more diverse datasets~\cite{merler2019diversity} and embracing positionality in data labeling practices~\cite{10.1145/3392866}. 

Natural language processing (NPL) is another area where gender has become a important issue. Many NPL algorithms have been found to amplify the gender stereotypes and biases that are present in dataset generated from human communications. Sun et. al.\cite{sun2019mitigating} summarized how these biases have been transferred to downstream applications such as machine translation and speech recognition, as well as different methods proposed to mitigate these biases from an algorithm standpoint.

\subsection{Actively supporting transgender individuals need from HCI}
Many transgender individuals face unique challenges due to their gender identity. On top of the discrimination and harassment commonly seen toward the LGBTQ+ community in general, many transgender individuals have to undergo medical procedures to obtain the correct physicality. 

HCI works targeting transgender individuals currently primarily focus on ensuring their safety. In \cite{SafeSpaces} the authors discussed technology design that can help build a safe space for transgender individuals. Haimson et. al.~\cite{trans_time} also explored a trans-focused social media site that allows users to safely share their transition experience. 

While constructing a safe environment for transgender individuals is doubtlessly crucial, directly addressing the needs of transgender individuals during their transition is no less important. Zhang et. al.~\cite{voice_training} detailed guidelines to develop a voice training application that assists individuals to alter their voices (speaking pitch), and especially how it should respect the diverse goals set by transgender individuals themselves. Our work aligns with this research direction, and we consider it important to demonstrate that we can also address the specific needs of transgender individuals to avoid framing them only as a population to be protected.  

\subsection{AGR System}
AGR refers to various computational methods that identify an individual’s gender from images, video, and/ or audio~\cite{chiagr2018}. Given an image, current AGR systems can ``gender", that is, to assign a label of either man or woman within milliseconds. Currently, AGR systems are extremely accessible, with Amazon, Microsoft, and other companies offering their own version of image tagging services in which AGR is embedded. These systems often utilize deep learning, a technique in which an algorithm ``learns" using a large amount of data (image and label)~\cite{LH:CVPRw15:age}. 
Datasets used to train AGR systems, such as the ``Diversity in Faces" dataset released by IBM, contains gender labels of only male and female~\cite{merler2019diversity} and often document no effort to ensure transgender individuals are represented.
Unsurprisingly, this leads to a performance degradation for AGR system on transgender individuals, averaging 70.5\% and 87.3\% for trans men and women respectively, compared to cis men and women averaging 97.6\% and 98.3\% respectively~\cite{scheuerman2019computers}. 

In their interview study with transgender individuals in the United States, Hamidi~\etal~\cite{chiagr2018} asked participants about their reactions regarding potential uses of an AGR system. Although they proposed scenarios that are ``not overtly positive or negative (i.e./ marketing, security, and human robot interaction)" , ``all (participants) questioned whether implementing it would offer any benefit to end users". As a result, the authors recommended future AGR system to offer consent, opt-out and incorporate gender diversity. 
We agree that the above recommendations are critical for any AGR use. Nonetheless, instead of people being solely subjected to AGR, we also need to explore how we might appropriate and redefine the ways we can use this type of technology proactively to support transgender people's needs.

\subsection{Virtual beauty makeup}
Virtual beauty makeup refers to any non-physical facial modification. Although many forms of virtual beauty makeup exist, such as light-projected cosmetic~\cite{bermano2017makeup}, one of the most wildly adopted form of virtual beauty makeover are smart phone virtual cosmetic makeup filters which apply transformations that mimic real cosmetic products (e.g., eye shadow and lipstick). 

With a large range of makeup effects, automatic makeup recommendation aim at providing a set of makeup that will match personal preference using questionnaires or on the basis of the facial features of a person. Scherbaum~\etal~\cite{scherbaum2011computer} ~ presented a method to extract facial makeup from 3D scans and automatically recommend makeup on the basis of facial feature similarity. However, to the best of our knowledge, none of the existing systems focus on makeup suggestions for transgender individuals.

%% file: 3-design_process.tex
\section{Design Process}
Our design process is grounded in a 3-year-long digital ethnography~\cite{doi:10.1177/0038038508094565} within Japanese trans communities, including online support group on Facebook, as well as on Twitter and other public forum sites. The goal of the ethnography was to learn about the impact of presenting as trans from different perspectives (medical, legal, social etc) and in multiple context (Japanese, English, and Chinese). 

With this ethnography, we observed that, especially in Japan, there is a huge demand for transgender makeup education, spanning across multiple media such as physical publications~\cite{9784860102678, 9784776911029}, YouTube channel~\footnote{https://www.youtube.com/channel/UCIrJi2rDr5RsmYOINg-La3Q}, and school~\footnote{https://www.satsukipon.com/otome}. More importantly, we discovered the wealth of practices around passing, including the desire to assess how well one passes, relying mostly on external judgement such as marketing research~\footnote{https://ameblo.jp/otomejuku-trans/entry-12369445400.html}. We also witnessed how transgender individuals in Japan often face difficulties when attempting to learn how to wear makeup. This phenomenon spans multiple social layers and starts with simple actions such as obtaining cosmetic products. Many transgender women consider themselves out of place in cosmetic shops or reported being afraid of being outed or discriminated. For those who live closeted with their families, it is even more complicated. Not only do they need to acquire cosmetic products stealthily, which rules out online shopping as the shipping package might expose them, but they also need to secure and apply cosmetics without being found. Afterwards, it could also be difficult to obtain feedback regarding their makeup if they lived in an unfriendly environment. All these factors negatively affect a potentially smooth makeup learning experience for transgender people in Japan. 

Therefore, this ethnography led us to identify the need for dedicated tools to support this activity and the potential benefit of a virtual makeup system that focuses on passing. We identified two different ways a virtual makeup system might assist transgender persons as they explore how different makeup may help them pass. The first one is to provide feedback as to how much a specific makeup might assist or hinder their ability to pass. The second one is to provide makeup recommendations that try to maximize the ability to pass. 

When designing for these two goals, we used the working hypothesis that both endeavors might potentially benefit from AGR systems if we change our perception of these systems and postulate that what they actually do is to evaluate passing. In doing so, one of our goals is to see if AGR can be subverted in some way for the trans-community despite all of their now well-documented limits. In that regard, Turkopticon~\cite{irani2013turkopticon} is a good example of how researchers might try to reuse existing questionable techniques to subvert larger systems. Before developing a fully functioning tool that could be released publicly, we want to use our first prototype as an additional formative stage to reveal transgenders’ personal perspectives and exploring with them the issues that need to be considered in the design of such tools~\cite{boehner2007hci}. We also think that this will help provide additional context for a critical debate around AGR.

To complement our digital ethnography, we also had two interviews with a doctor working in a mental clinic specializing in gender development in Tokyo. We had our first interview during the ideation stage. At this point, we presented the findings from our ethnography and focused on the proposed features, ensuring that our system might align with the need of participants. We also showed a non-functional prototype and discussed with the doctor whether he encountered the need of each individual feature (i.e., scoring and recommendation) during his daily practice. In the second meeting, this time with a working prototype, we checked about participants’ safety, in terms of system use, study procedure and questions to ask. These two interviews helped us confirm the potential benefit of developing a virtual makeup system that supports trans-specific needs, especially enabling a safe and private place for exploring how different makeup styles affects their passing.
We later decided to invite the doctor to join us as the third author after all the interviews had concluded, given all the valuable discussions we had prior to the study as well as further discussions that helped us understand the data.

%% file: 4-implementation.tex
\section{System Design and Implementation}
Our design decisions were grounded in the ethnographic work.
Following the two aforementioned design goals, our virtual makeup system, \emph{Flying Colors}, lets users load a picture of their own, choose their desired passing gender (man or woman) and explore a set of virtual makeup on the picture. Users can apply and combine virtual makeup components (foundation, hairdye, highlight, contour, cheek, lipstick, eyebrow, eyelight and eyeshadow) by selecting a set of predefined \textit{shapes} and \textit{colors} for each of them using buttons. An additional slider lets users modify the \textit{intensity} (i.e., the amount) of makeup for each component. On the basis of the virtual makeup applied on the picture, the system provides a makeup feedback by computing a score that indicates the effect of the makeup on the level of passing of the person (i.e.,\ how much does the makeup assist or hinder the user's ability to pass). This makeup feedback always starts from 0 when the users initially upload their image. When the makeup components are toggled or adjusted, virtual makeup components are rendered and the makeup feedback is updated accordingly. In addition, the system also provides five automatic visual makeup recommendations that are based on the makeup feedback. The system runs locally and the pictures are never shared with any third party.

\subsection{Virtual makeup}\label{subsec:makeup_ops}
We initially planned to re-purpose commercial virtual makeup applications, such as SNOW~\footnote{\protect\url{https://play.google.com/store/apps/details?id=com.campmobile.snow&hl=en&gl=US}} and YouCam~\footnote{\protect\url{https://play.google.com/store/apps/details?id=com.cyberlink.youperfect&hl=en&gl=US}} to include dedicated passing support features. In doing so, we would have had access to a large and accurate range of virtual makeup provided by the application. 
However, extending existing mobile app proved to be technically challenging. Therefore, we created a similar desktop application from scratch. 
We chose to design our tool for no specific gender identity. To do so, we borrowed make up templates from a commercial Korean application (SNOW) that does not strictly focus on makeup for women and is hugely popular in Japan with more than 22,000 ratings on Apple iOS App store. Closely following the commercial applications, we implemented a small set of makeup components to mimic existing physical cosmetic products. They include foundation, hair dye, lipstick, eye shadow, cheek, highlighting, and shadowing. We did our best to select make-up templates that work for Asian men (e.g. cheek contouring) and women (e.g. chin contouring) respectively, by looking at trans focused makeup websites. 

The virtual makeup components are applied in one of the two methods, segmentation based and 3D facial model based. Both of them are customizable, and the user adjusts the color and intensity continuously for each component independently.
The segmentation based method is used for components where the facial feature is visibly well defined in the picture (foundation, hair dye, lipstick, and eyebrow). 
We use the 3D facial model based method for the components that only fill a partial area inside of a segment (eye shadow, eye highlight, cheek, highlighting, and contouring). The 3D facial model based method allows adjustment of shape on top of color and intensity by selecting one of the makeup design templates that change the shape of the component. We recreated the templates following the ones available in commercial applications. 

The segmentation based method is carried out by first applying facial part segmentation on the user's face using the method proposed by Nirkin~\etal~\cite{nirkin2018_faceswap}. The resulting segmentation maps contain the pixel-level information of the position of facial parts (e.g., hair or eyebrows). We then apply linear interpolation between the input images and the target makeup color in HSV color space.

The 3D facial model-based makeup is applied by first reconstructing a 3D facial mesh from the image using a 3D Morphable Model~\cite{bfm09}. After the mesh reconstruction, we render the design templates on top of the image. The makeup templates are ``masks'' with makeup designs such as eye shadow of a single color, drawn on top of the UV map of the morphable model. 

\subsection{Makeup feedback}\label{subsec:lop}
We defined the \textit{makeup feedback} as the change in probability of an image being classified as a specific gender after applying the virtual makeup. 
Because we are specifically targeting transgender individuals, we chose to use a standard binary image classification neural network as a starting point. The network outputs $2$ probabilities, $p_f(I)$ and $p_m(I)$, given any input facial image $I$. $p_f(I)$ and $p_m(I)$ represent the probability of the person in the image being classified as female and male respectively. It is worth noting that $p_f(I)$ and $p_m(I)$ are computed independently; hence, they do not always sum to 1. 
Using the classification, we define the makeup feedback as $p'_x(I) - p_x(I)$, where $p_x(I)$ is the probability of the input picture being classified as gender $x$, and $p'_x(I)$ is the probability of the input picture being classified as gender $x$ after applying virtual makeup. For example if the objective is to pass as female, then the makeup feedback is computed as $p'_f(I) - p_f(I)$ and $p_m(I)$ is discarded. In the current version the desired passing gender is inputted using command line when launching the system. The makeup feedback shows the relative improvement of passing. 

While the idea of providing a numerical score led to much discussion within the research team because it might lead to the false impression that this feedback is ``accurate and rational,'' we still decided to display a numerical feedback as it seemed to us that other representations would be hiding the original nature of the score provided by the AGR system. In our implementation, the makeup feedback always starts at 0, and when makeup is applied, it can take any value between $-1$ and $1$.  

To train the gender classifier, we chose to use all images labeled as ``Asian'' in the UTKFace dataset~\cite{zhang2017age} because this dataset covers a wide range of photo conditions and contains label with race and gender. Another candidate dataset was the Chicago face dataset, which offered higher image quality but was rejected as it was captured in a studio setting which might affect performance given our specific use case. We chose to restrict our dataset to Asian faces because all the study participants are Asians, and facial structure has a fundamental effect on the makeup. However, we acknowledge that this classification is reductive as it does not accommodate people who have part-Asian ethnicity for example.
According to its authors, the UTKFace dataset is labeled using the DEX algorithm~\cite{DEX-ICCVW-2015} (trained on the IMDB-Wiki dataset) and verified by a human annotator. Following the inquiry of Scheuerman~\etal~\cite{10.1145/3392866} into how gender is constructed in image databases, we acknowledge that this classifier has severe bias in terms of both data collection and potential annotation error. Although data bias will most certainly affect the accuracy for the classification for people of less represented groups~\cite{buolamwini2018gender}, we are not using it to determine gender but instead to provide feedback about the effectiveness of makeup for passing. We think that in our case, these biases have a relatively limited impact on our goal.

\subsection{Makeup Recommendation}\label{subsec:makeup_optim}
To generate makeup recommendations, we define the \textit{desired makeup} to be makeup that increases the user's passing. In other words, we recommend makeup styles that maximize the makeup feedback. Since the operation of selecting a template is discrete, we cannot directly optimize it using continuous methods such as L-BFGS~\cite{liu1989lbfgs}. For a component with shape templates, we first perform a 2D grid search on color and intensity for each template individually. The top five highest rated templates are selected and then jointly optimized for color and intensity using random optimization. We present the top five highest scored combinations in the user interface. Users can explore these combinations using the ``Recommend'' button.

\begin{figure*}[ht]
    \centering
    \includegraphics[width=0.8\linewidth]{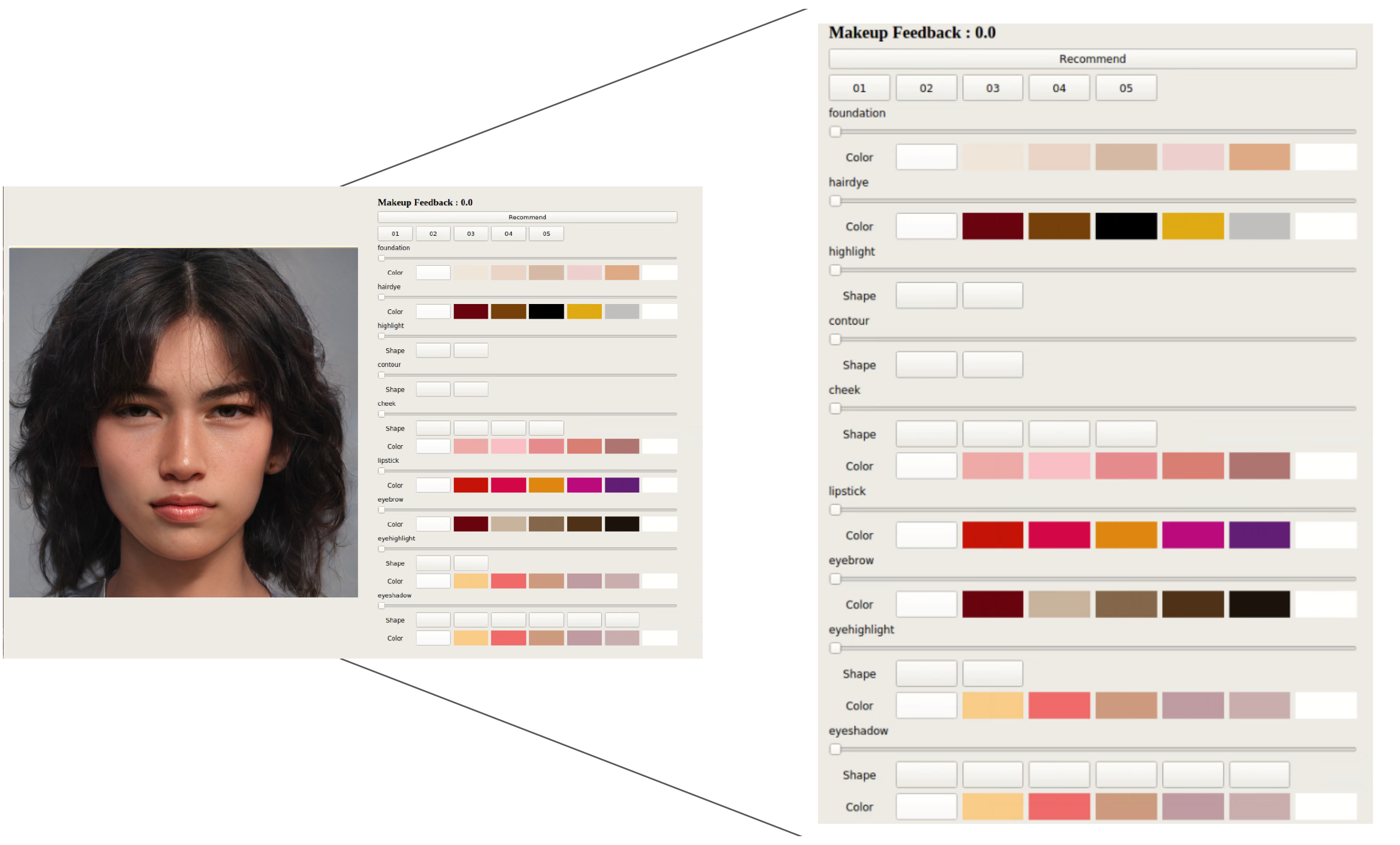}
    \caption{
      Screenshot of the user interface. On the left side, the user's front-facing image is displayed. On the right side, the  \textit{Makeup feedback} of the user (Sec~\ref{subsec:lop}) is displayed at the top. The \textit{Recommend} button when clicked, computes the top five highest passing score makeup, which are displayed when clicking on buttons \textit{01} to \textit{05} below (Sec~\ref{subsec:makeup_optim}). Each makeup operation (Sec~\ref{subsec:makeup_ops}) is listed with the makeup component name, a slider to adjust the intensity of the component, shape (i.e., design templates) and color buttons, which are available when the operation supports multiple shape designs and color variations respectively. The leftmost color button (white) triggers a color palette to let users select their preferred color.
    }
    \Description{Our user interface shows am frontal face mage of the user on the left side. On the right side is from the top, makeup feedback (a numerical number), recommend buttons, and a list of different makeup components as well on the right. Each of the makeup components (foundation, hair-dye, highlight, contour, shape, cheek, lipstick, eyebrow, eye highlight, eye shadow) as a slider that allows the user to adjust the intensity of the components. Some components (foundation, hair-dye, cheek, lipstick, eyebrow, eye highlight, and eye shadow) comes with color buttons, which allows the user to either use one of the five preset colors, or the user can pick a custom color using the right most button which opens a color palette. Some components (highlight, contour, cheek, eye highlight, and eye shadow) come with different design templates that allows the user to adjust the shape of the component.
    }
    \label{fig:network_arch}
\end{figure*}

%% file: 5-user_study.tex
\section{Study}
\label{sec:user_study}
Our goal with this study is twofold. We first want to document and understand what are transgender experiences with passing, with virtual makeup systems as well as AGR systems in Japan. We also want to explore how they interact with and interpret our virtual makeup system.

\subsection{Participants}
We recruited $15$ transgender participants with the help of the local gender clinic we interviewed. We put a poster inside the clinic indicating that ``we have developed a virtual makeup system that provides recommendation to help you look more feminine or masculine'' and that we are looking for feedback. Participants had a chat with the attending physician before enrolling in the study. They were at very different stages in their journey as trans and had very diverse technical background regarding machine learning. In Table 1, participants who have a basic understanding of machine learning are classified as intermediate, and participants who work with machine learning in a professional setting are classified as expert. 

\begin{table*}
\centering
\small
\begin{tabular}{|c|c|c|c|c|}
\hline
\textbf{ID} & \textbf{Gender identity}     & \textbf{Makeup habit} & \textbf{Age group} & \textbf{Technical background} \\ \hline
P1                           & Trans woman                                   & Every day                               & 31$\sim$40                          & Intermediate                                   \\ \hline
P2                           & Trans woman                                   & Every day                               & 21$\sim$30                          & Intermediate                                   \\ \hline
P3                           & Trans woman                                   & Every day                               & 21$\sim$30                          & Intermediate                                   \\ \hline
P4                           & Trans masculine, non-binary                   & Special occasions only                 & 21$\sim$30                          & None                                           \\ \hline
P5                           & Trans woman                                   & Every day                               & 41$\sim$50                          & Intermediate                                   \\ \hline
P6                           & Trans woman                                   & Every day                               & 21$\sim$30                          & None                                           \\ \hline
P7                           & Trans woman                                   & Every day                               & 21$\sim$30                          & Intermediate                                   \\ \hline
P8                           & Trans woman                                   & Every day                               & 21$\sim$30                          & None                                           \\ \hline
P9                           & Trans man                                     & Special occasions only                 & 21$\sim$30                          & None                                           \\ \hline
P10                          & Trans man                                     & Special occasions only                 & 21$\sim$30                          & None                                           \\ \hline
P11                          & Trans man                                     & None                                   & 41$\sim$50                          & Intermediate                                   \\ \hline
P12                          & Trans feminine                                & Only at home                           & 21$\sim$30                          & Expert                                         \\ \hline
P13                          & Trans woman, genderqueer                      & Every day                               & 31$\sim$40                          & None                                           \\ \hline
P14                          & Trans feminine                                & Every day                               & 31$\sim$40                          & None                                           \\ \hline
P15                          & Trans female                                  & Every day                               & 31$\sim$40                          & Intermediate                                           \\ \hline
\end{tabular}
\normalsize
\caption{Table of the participants, including details about their gender identity, makeup habits and technical background regarding machine learning.}
\end{table*}

\subsection{Protocol}
\label{subsec:protocol}
The study was divided into three sections: an initial introduction and interview, an exploratory session with the system and a final discussion session focusing on participants' experience with the tool. The following protocol was first discussed with and approved by the local clinic doctor and then approved by the university’s review board. 
Participants received the description of the study prior the study and signed a consent form. We first asked the participants to fill out a short questionnaire about their demographics and trans-identity as well as their past and current makeup habits and experience (if any). Based on their response, we followed up with a semi-structured interview to understand their makeup habits, their experience with virtual makeup systems, and their experience of passing with people and with machines. This introduction section lasted around 45 to 60 minutes. 

Following this initial interview, we demonstrated our system with an image from the dataset. We then input participant face image as well as their desired passing gender. We asked participants to re-create their own everyday makeup or their last makeup if they do not wear makeup regularly. After that, we asked participants what they thought about the makeup feedback. We repeated the same process with the goal of exploring new makeup styles instead of recreating personal makeup styles. We showed the system recommendations and asked participants what they thought about the recommendation or indicate what aspects of the system they would like to change or see us improve. In the third part of the study, we asked participants about their opinion about this particular use of AGR system, especially if it triggered any negative or discriminatory feelings, and more general questions regarding the system design. At this point of the study, we also explained and discussed the mechanism behind the makeup feedback. Finally, the participants answered a short questionnaire to evaluate the system. Participants received a ¥4000 (approximately USD\$40) gift card at the end of the study and we deleted the facial image(s) used for the study.

\subsection{Data Analysis}
The first author conducted and transcribed all the interviews. All interviews, except for P9 and P12, were conducted in Japanese. They were directly translated into English during the transcription process. Interviews for P9 and P12 were conducted in a mix of Japanese and English. Japanese portions of the interview were translated and English portions were directly transcribed. Note that all quotes are not translated word-for-word but reworded to ensure readability. We made our best to keep the participants original meaning intact. 

We analysed the interviews using the approach to thematic analysis developed by Clarke and Braun \cite{clarke_thematic_2014}. Two of the authors regularly met to discuss the data and the coding to ensure that the themes for each of the research questions were coherent, comprehensive, and reflective of the actual data. When needed, we presented direct quotes from the participants together with underlying social expectation and assumptions of gender role in the Japanese society to provide context for readers unfamiliar with the Japanese culture. We agree with Clarke and Braun that even single occurrences of a theme can be important. We provide counts when appropriate to help readers contextualize the feedback we received but these counts should not be interpreted beyond the scope of this study.

%% file: 6-findings_makeup.tex
\section{Findings - makeup and passing}
We first present findings about participants' experience with passing and makeup in Japan. We then report on their experience with existing virtual makeup systems as well as their strategies for evaluating their passing. In Section~\ref{sec:findings-sys} we analyze and discuss participants' feedback and perceptions of our makeup system.

\subsection{Motivations for passing}
In our pre-interview questionnaire, 13 participants indicated that passing is very important for them (6 or 7 out of 7-point Likert scale), 3 indicated that passing is quite important for them (4 and 5 out of 7) and 1 indicated that passing is not very important to them (3 out of 7). In our case, participants especially wanted to pass in public and work settings. All 13 participants explained that avoiding discrimination was one of their major motivation of passing. P11, for example, often get ``a weird look'' in public areas especially in bathrooms and changing rooms. P3 thinks passing is an effective measure to avoid people who are passively transphobic. ``I think passing is going to help me reduce the number of times I encounter those people [who do not actively seek out transgender individuals but start bothering me if they notice that I am trans], [...] so it's just about not being noticed''. 

One of the most important places for the participants to pass is at work. Many expressed frustration and resentment about the fact that bosses and coworkers who knew them before they started medical transition will often never accept them for who they are. Maintaining a functional social life and career has become a top priority for them. P1 stated that her motivation to pass is to prevent ``my coworker and bosses [to] take precaution against me or think of me as a weird person''. This might be due to the fact that $70\%$ of the LGBT community have seen or experienced harassment in workplace in Japan and little legal protection is offered to the community (i.e., no national anti-discrimination law for sexual orientation and gender identity). Even worse, reflecting on the workplace P1 worked for when she began her medical transition, she ``was treated as a male and was told to behave like on''.
Similarly, P4, who work as a personal trainer, expressed how passing is essential for their work, ``actually I prefer X-gender or non-binary (as my gender identity), but for work, ... there is the changing room etc. ... a new instructor who joined was MtF had caused quite some problems. So I present as male at work even thought I don't consider myself one''.


\subsection{Experience with makeup}
Although voice was considered the most important factor for passing by the majority of participants, they also thought that makeup was an important aspect for passing. Most of the participants in this study showed interest in using makeup. Of the 15 participants, 8 wear makeup regularly and 13 have used makeup in private settings. Among them, 8 mentioned that they prioritize passing when wearing makeup and 5 prioritize self-expression instead. 

\subsubsection{Learning about makeup}
10 participants proactively looked for new makeup products or methods to improve their own makeup skills, such as watching makeup tutorials on YouTube or following social media influencers who produce cosmetic-related content on platforms such as Twitter and Instagram. It is noticeable that while participants learn about makeup products from all kinds of resources, many of them learn to apply those products specifically from transgender and crossdresser-oriented sources primarily. P13 expressed her view on cisgender-oriented makeup tutorials: ``I feel like they could be useful for reference. But if you apply this [cis-woman makeup] makeup to a [biologically] male's face as is, it will emphasize the masculine part of the face, this is especially true for Japanese people [given their facial features]''.
Many participants expressed their difficulty talking to sales representatives at cosmetic counters, and some relied on events that specifically give out makeup advice for LGBTQ, for example P3 who shared that: ``There was a gender-free makeup/fashion event where I met a friend from (a cosmetic company)''.

\subsubsection{Obtaining cosmetic products}
7 participants also pointed out that there are difficulties surrounding obtaining cosmetic products in store with the fear of being identified as a transgender person. Some of them also felt out of place, often citing their lack of confidence in their appearance as an important factor. As counter measure, P8 for example shopped with friends or partners to reduce the chance of being outed and to prevent potential awkward situations. 
Some justified their purchases to the sales clerk by claiming it is for stage performance (P1) or is a gift (P3).
Participants who do not wish to interact with the sales clerks often resorted to online shops or generic drug stores for their makeup needs. This makes it difficult for them to test products or obtain professional advice before purchase.

\subsubsection{Applying makeup and obtaining feedback}
After obtaining cosmetic products, many participants still often lack opportunities to apply it or obtain feedback. In many cases, it is due to the lack of family (parents or partners) support. 
Some participants started their transition at a younger age. In P7 case, she could not apply makeup at home due to her parent's objection, and could only do so after working at a `Newhalf bar', a bar with trans-woman hostesses. P6 and P14 did not have fully supportive partners and were not always able to have makeup related discussions. In P14's case, her partner said that she wanted ``nothing feminine in [their] house''.

Social acceptance plays a huge role for men wearing makeup. P9 wears makeup when he is cosplaying (i.e., dressing up as fictional characters) and would like to be able to do so when he goes to job interviews in order to be presentable. However, he was instructed by a government employment service center staff not to wear makeup since the interviewer ``might get suspicious if a man's skin look too clean''. The social expectation of `men should not wear makeup' restrained participants from adopting makeup.

\subsection{Experience with existing virtual beauty systems}
Out of all the 15 participants, 10 have tried modifying their appearances with virtual systems. Some participants highlighted how they used this kind of systems for fun, to experiment with different makeup styles. A few participants also explained how they creatively use existing virtual beauty systems. For example, P8 uses SNOW ``to decide on the general direction of the surgery'' before talking to the doctor. P10 used FaceApp to make his face more masculine before sending a couple picture to his in-laws who believed he was cis-male. These applications are examples of transgender individuals proactively repurposing tools for their specific needs.

One of the key limitations of current virtual makeup systems is the fact that they are often designed for cis-women. All trans-woman participants highlighted the fact that makeup for trans-women is very different from that for traditional ciswomen, and simply applying templates not designed for them often result in a sub-optimal look. ``Woman makeup is more about enhancing features. But for us it's more about hiding them'' (P1). Similarly P5 explained: ``If I apply female makeup directly, I will look more masculine, so I have been applying makeup in a reverse style''. Therefore, some of the participants were not satisfied with existing virtual makeup systems: ``I think the makeup they provide don’t fit my facial feature and don't look good on me'' (P2).

\subsection{Evaluating how one passes}
\label{subsec:eval_passing}
\subsubsection{Evaluating how one passes with people}
In our initial questionnaire, of the 15 participants, 10 responded that they had used non-technology-based ways of evaluating how well they pass. Direct approaches include directly asking friends or strangers, for example, asking for bathroom location in shopping malls where women and men's rooms are on different floors. Going to stores is also a common strategy, P5 explained that ``for some shops, on their receipt, like [name of a chain restaurant] and some convenient stores as well, they [the employees] have to gender you for marketing''. However, many also noted that this method can be unreliable sometimes. P1 thought the method is ``rather ambiguous, there are cashiers who press the male [button on the machine] no matter who comes'', so instead she ``talks to strangers on online chat roulette, there are people who want to talk to girls, and they will leave as soon as they realize that they are not talking to a woman''. Passive approaches, such as observing others' reaction are also practiced among some participants, P8 explained that ``I work in a place where I handle a lot of kids, I usually approach a kid without telling them my gender, and I see which gender they think I am''.

\subsubsection{Evaluating how one passes with AGR}
6 participants had already used software systems to voluntarily evaluate how well they pass, prior to the study. For example, more tech-savvy users (P1, P5, P12, and P15) used Web services (e.g., social media sites that estimate the user's gender using his/her profile picture) or APIs (e.g., Microsoft Face API) that provide an AGR function and ask the system to gender themselves.
These services provide either a binary prediction (i.e., male or female) or a prediction with probabilities of the person in the picture being male or female. Using a software-based system for evaluating how well they pass meant that participants could experiment with their appearances. For example, P8 explained that she ``change[s] the way she takes pictures such that [she] can approach [high woman probability score]''. Participants also appropriated systems originally designed for other purposes. For example P1 discussed using an app that finds ``a celebrity who looks like me and see whether that person is male or female''. 

Although all of the above activities were carried out by the participant alone and willingly, some participants also experienced public evaluations with AGR-enabled systems often deployed publicly for marketing purpose. P15 mentioned a digital signage she found in an exhibition that used AGR to recommend product ``dedicated" for different gender and age groups. She explained that AGR tech was interesting as it showed different results depending on different viewing angles, which aligns with her own personal feeling. Although she acknowledged that the technology could be useful for marketing and could be validating to her gender identity, at the same time, she was concerned about the involuntarily outing that public deployment of AGR could potentially cause. Similarly, P9 found a vending machine equipped with camera to automatically recommend drinks according to the user's gender. He was excited to use the machine at first but was misgendered and let down. He disagreed with the deployment of AGR technology in this fashion because `there are much more diversity than AI can deal with'. Even in a private setting, not all uses of AGR-enabled software systems were intentional. P5 explained that Google Photos automatically creates albums of each individual from the user picture library and that she was classified as two different persons, before and after starting hormones replacement therapy. She explained that she actually preferred this behavior: ``If the system can tell that I am the same person as when I was a male, then a security camera could tell my assigned gender, which is very threatening and scary'' (P5). It is worth pointing out that participants negative feelings echo the ones reported by Hamidi et. al.~\cite{chiagr2018}, thought the negative attitude from our participants are mostly focusing on being subjected to AGR system involuntarily, and hold a more neutral to positive attitude to the technology itself.

We also need to mention that the motivations behind the desire to evaluate how one passes are very complex. Aside from understanding their personal progress on medical transition/makeup skills, participants often hinted at or pointed out seeking potential validation as one of the reasons. For example, P12 said that she did it feeling that ``she is playing a weird game'', that she ``feels sad or unhappy'' about her makeup skills if the result does not align with her desire, and ``validating'' if she could pass without makeup. This ``thrill'' of seeking validation and opinion from a machine, while understanding that it might not align with their own feeling can be seen in many other participants as well. P10 expressed interest in engaging in evaluation, yet had never proactively evaluated himself, citing fear as the main reason since he is trying his best to pass and could not handle the potential scenario in which he would be told that he is not passing.

%% file: 7-findings_system.tex
\section{Interacting with the virtual makeup system}\label{sec:findings-sys}
Regarding the virtual makeup system in general, we received overall neutral to positive feedback. We present below detailed feedback about the two key features of the system: the makeup feedback and the recommendation feature.

\subsection{Interacting with the feedback system}
\subsubsection{Positive perceptions of the Makeup feedback feature}
We received overall neutral to positive feedback from the participants regarding the makeup feedback system. Even if our goal was not to evaluate the algorithm in itself, we wanted to observe how participants' perception might align or not with the system and how they might react to this alignment or misalignment. When asked in the post-hoc questionnaire whether they agreed with the makeup feedback when trying out makeup, participants responded an average of 4.8 out of 7. 

Several participants found that the makeup feedback could be validating regarding their makeup skills. For example, P4 thought they could use it to reassure themselves about their makeup: ``Before heading out to a special occasion when I want to show up nicely, I can go through it just to make sure." While not agreeing with the feedback entirely, P12 indicated that the system can be used as a reference when trying out new makeup styles that allows them to compare different makeup styles easily. Interestingly, P9 praised the system for demonstrating that gender is non-binary: ``by presenting it as a score, it doesn't present gender as a binary concept. [...] I can adjust my own expectation." Transgender individuals do not always have identical expectations regarding their transition depending on their transition stage. This is often further affected by their biological and financial situation. In that case, having a numerical feedback allow users to set their own passing goal.

\subsubsection{Unexpected use-cases}
A few participants proposed different and unexpected use cases for the system. For example, P6 explained that the only use she could think of for this feature would be ``just keeping track of how well I pass [over time], and revisit the good ones for reference." Because he does not wear makeup regularly, P10 thought that the makeup feedback would help him stay focused on his goal of passing: ``if the system can show me the goal of my makeup it would be very useful. There could be times when I overdo my makeup etc, I think we can prevent that by showing a goal". P2 also proposed to validate her own makeup by `comparing my before and after makeup'.

A few participants felt that computing the feedback from a single picture could be limitating. P10 explained: ``I feel like it is not so reliable, it is probably better to have a video''. Interestingly, the same participant also perceived this imperfection as a self-protection mechanism. He noted that ``if it is a machine, even if the score is low, I feel like I can convince myself that it is just a machine, I can fault the machine instead of myself. [...] On the other hand, if it is a scenario where we ask 100 people to answer a questionnaire [and the score turns out to be low] that will be quite disheartening''. When comparing the system with the experience of being misgendered in their real life, P3 explained that to her ``it's much worse to be told [that you are not passing] in real life''.

Participants for whom passing is either externally motivated or not a priority at all did not find the feedback useful. For P4 whose sole motivation for passing is to avoid discrimination and would otherwise not want to pass, the makeup feedback was not useful as they simply ``don't really care [at a personal level outside work]". However, even participants who did not find the feedback useful for themselves thought that it would be useful for someone they know. 

\subsubsection{Issues with the Makeup feedback}
No participants reported feeling uncomfortable when prompted about their experience with the makeup feedback feature. Nonetheless, a majority of participants identified potential issues with the makeup feedback in its current form. 
P9 thought the makeup feedback would discourage users from exploring new makeup styles that score low with the system but also thought it might have more to do with ``anatomy and makeup trend" than the system.
Making a connection with his background in agriculture, P11 was concerned regarding how the system would be perceived by non-binary people ``there will always be people in the middle, mentally, physically ... And hearing from you that, we are at the point where we can clearly classify man and woman using technology sounds quite scary''.   
While exploring the system, P7 jokingly acknowledged the potential damage the system might induce: ``I think [the score] is a good feature, but I don't think it will make me happy. It's like a battle within myself. If the machine displays a popup that tells me that my sense of makeup is bad, I will probably get mad''. They also mentioned the potential addictive nature of such a system, ``It is like watching a horror movie, you know the risk that the score could be high or low."

\subsubsection{Makeup feedback vs Passing feedback}
While the makeup feedback $p'_x(I) - p_x(I)$ is the only feedback we provide to the user, an early participant proposed and we discussed with the other participants regarding a hypothetical variant of the makeup feedback. We named it \textit{passing feedback}, defined as $p_x(I)$, an absolute value that describes the person's ability to pass to the classifier.
When we explained this distinction to the participants, they all agreed with our decision to display only the makeup feedback. For P1, providing only a makeup feedback is ``the more peaceful, less provoking, way to go". P5 considered that a raw passing feedback would be a discrimination towards transgender individuals as ``if the starting score isn't 0, say it's plus or minus, then I think in this case the score will be more favorable to cis-woman". However, while all participants favored a makeup feedback and acknowledged the controversy and harm, 7 showed interest in knowing the raw passing feedback for themselves. P8 thought the passing feedback would be beneficial for herself, ``to me at least, knowing the raw passing feedback is useful to decide the makeup direction for being who I am."
P1, who also considered the makeup feedback to be less controversial still thought that ``people would naturally have different level of passing, it would be useful to know what is their passing score at the beginning." With professional knowledge of AGR algorithms, P12 explained how understanding the algorithm behind could be the key to the passing score acceptance, ``it would be good to just show the difference (makeup feedback) for public use but as for myself, I would like to see the raw (passing) feedback, because even if it does not follow my ideal, I can still understand it because I know how it works, and maybe that will not make me feel so sad. But maybe for others if they don't know how it works, maybe it feels like another stone in their hearts."

\subsection{Interacting with the recommendation system}
The current recommendation system provides five different sets of makeup that try to maximize the user's level of passing. With regards to these recommendations, we received mixed feedback from participants. 
When asked about how useful this system was for exploring new makeup styles, the 15 participants rated the system on average of 5.3 out of 7 on the Likert scale. This rating echoes the feedback of participants who were not fully convinced by the current implementation but were nonetheless positive about the idea of having a virtual makeup system dedicated for passing. The two key limiting functions were the lack of customization features and limited realism.

For P13, for example, a recommendation system was warmly welcomed: ``I will just stick to the recommendation and will not buy anything else. When I tried my personal preference, my score went down. Passing is the sole purpose of me wearing makeup, I don't see any reason to do it differently." She also commented : ``Also usually I don't really use eye-shadow, this system really encourages me to try out different patterns and color." P3 also explained that ``it's a great chance to know what color makes me look more feminine. Knowing what color fits me will help me find products afterward".

Participants also had a lot of ideas about how this type of recommendation system could be improved. Currently, each of the makeup components are jointly optimized using random optimization, so there is no guarantee that all makeup components will fit under the a particular style. P1 suggested ``it would also be nice to be able to pick a style of makeup like pretty and cute etc." Similarly, participants often had preference for certain cosmetic products. P3 expressed her desire to be able to preset ``a cosmetic product choice that I simply won't change anything about, its better to able to incorporate those and build upon it." The makeup recommendation being computed with a single picture led to other concerns. P10 questioned about ``this makeup [pointing to a recommendation], how does it look from the side, there are also cases when makeup and clothes don't match. I would want to see the total effect instead of just the makeup."

However, the recommendation system currently do not optimize based on semantic information, so makeup recommendations do not always look consistent (i.e., components could have drastically different colours that did not fit together). In addition, makeup recommendations are not semantically associated with any makeup style (e.g., kawaii or cool beauty) commonly found in popular culture, which many participants preferred. 

\subsection{Improving the system}
When asked about how useful this system was for exploring new makeup styles, the 15 participants rated the system on average of 5.3 out of 7 on the Likert scale. This rating echoes the feedback of participants who were not fully convinced by the current implementation but were nonetheless positive about the idea of having a virtual makeup system dedicated for passing. The two key limiting functions were the lack of customization features and limited realism. Participants proposed numerous ideas to improve the system.

\subsubsection{More customization options}
Beyond the two core features of the tools, participants proposed numerous ideas to improve the system.
In particular, participants showed interest in having more customization options. Our system provides several templates for specific makeup operations. During the interviews we received requests to add more customization options, such as editing the shape of the eyebrow or changing the shape of the lips with lipstick. Additional functions from other commercial software systems such as pimple removal and skin smoothing were also requested. We had based our makeup templates on existing commercial applications that do not strictly focus on women, but they still proved to be quite limiting, especially for trans-men. P11 wanted to be able to explore more options : ``more customized for FtM, like [...] attachable mustache, eyeglasses, and hairstyle".

\subsection{Being subjected to an AGR system}
Beyond exploring our system, our goal was also to explore with participants their opinion regarding interacting with an AGR based system. Given the nature of our tool and how we recruited participants, participants who proactively work to pass in their daily lives may possibly be over-represented. While it is certainly not possible to generalize this finding, in this study participants viewed our system merely as a tool to assist them with their goal. P3 compared our system with being gendered in real life, ``I don't think this is discrimination. It is much worse to be told [that you are not passing] in real life." Participants who do not wear makeup regularly generally found our system less useful, yet, in our study, none of them viewed it as discriminatory. P11 explained ``I think this is only a problem if you force this onto someone. But if you are giving people a choice, I think it is fine. For transgender people who decide they want to go through the transition, many of them care a lot about passing and they will use it. For those who don't care about what other people think about them, they can simply not use it." 

Interestingly, beyond makeup improvements, P12 also wanted to push further the use of machine learning: ``It would be better if you provide like a heat map [...] some visualization about the network... Which part is important, which part of your face contributes more to male or female''.

%% file: 8-discussion.tex
\section{Discussion}
\subsection{AGR as a potential ally}
Results from the study show that AGR based systems can potentially be leveraged as tools to support people who wish to pass.  

\subsubsection{The perks of AGR systems: Confidentiality, Consistency, and ``Blameability"}
Unlike methods relying on external human judgement that are currently explored by participants to evaluate how well they pass (see section.\ref{subsec:eval_passing}), AGR can provide complete confidentiality with no risk of letting other individuals know or impact the evaluation result. In that sense, AGR can provide a safe space for exploring makeup without any external consequences. The makeup feedback mechanism strictly focuses on the effect of makeup and removes the additional stress that can come from a global passing evaluation. In contrast, it may be more difficult for human to disentangle the effect of makeup on passing.

Even if it has important limitations, once it is trained a computer algorithm will generally provide consistent results and, in our case, this can be considered as a desirable property. Human judgement often varies across individuals, or even within the same individual (i.e., mistakes or low-effort evaluation). In contrast, participants expect the system to produce a consistent evaluation that they can rely on to improve their makeup skill over time. AGR can potentially create a predictable and reassuring environment that minimizes stress. The makeup feedback score, because it is not binary, not only reveals that gender is not a binary concept but it also lets trans individuals establish their own standard and keep track of their progression, even if minimal. 

As mentioned by P10, AGR as a human-made artefact offers an additional sense of `blameability', the ability to selectively disregard the evaluation. By acknowledging the system as being imperfect, trans individuals can find it easier to ignore feedback that they do not agree with. Evaluations from actual humans, as subjective and arbitrary as they may be, cannot be disregarded as easily because they can always trigger real life consequences such as discrimination. How to design a `blameable' interactive system remains an open question that future work should explore.

\subsubsection{Appropriating AGR systems to support diverse (non)passing goals}
Our system is currently designed to provide assistance to maximise the passing score. However, it became evident during the study that focusing on this unique goal is very limiting as participants had more complex and nuanced desires than simply trying to optimize for a high score. Instead, participants saw other potential use cases for an AGR based makeup system. For example, they wanted to keep track of one's progress in terms of passing or to search for makeup that balances personal preference and passing score, for example by pre-selecting makeup components and adjusting others accordingly.  

Given our system technical implementation, we could also easily extend this tool to assist non-binary people or people who would prefer to explore makeup that explicitly do not conform with traditional gender norms. With AGR being a 2-class binary classifier that computes the probability of a certain input being male and female independently, even reusing the current implementation, a non-conforming goal can for example, be expressed as minimizing the probability of being recognized as male and female at the same time. Beyond our current implementation, developing datasets that specifically cater to transgender passing goals could also open doors for other use cases. Current data-sets always focus on binary notion of gender but could be relabeled using more gender labels so as to provide more passing goals for example.

\subsection{Limitations of an AGR based system}
Although our findings suggest that AGR-based systems could potentially provide assistance to support transgender exploration of makeup for passing, we must state that this can only be true under strict conditions. Privacy rules as well as revocable consent-based system design need to be enforced to ensure that such systems are only used when transgender people wish to use them. And even so, we still identified potential issues with our system and AGR systems in general. 

\subsubsection{AGR as a potential threat and ways to mitigate them, including privacy and effective consent concerns}
One of the recurrent themes in responses regarding system design, is that our system is designed for makeup exploration, which is often conducted in a safe space and hence is an acceptable context. Most participants explicitly explained that while they would be keen on knowing the passing feedback for themselves, they are uncomfortable sharing it with others.

In contrast, many publicly deployed AGR systems, such as surveillance cameras and the vending machine mentioned by P3, are always recording with no indication or consent requirement. Whether or how the data collected is protected is often unknown to the user. In particular, one of the threats is that this technology could potentially be used to expose transgender individuals' assigned gender. 
We also acknowledge the risk of AGR being exploited to disadvantage transgender people as a passing metric they would be required to comply with, such as employers using it as a metric for hiring or requiring employees to attain a certain `level of passing' at work. We believe that AGR can be beneficial to the community when deployed for their own needs but only if users voluntarily subject themselves to the system instead of the opposite way which has been the case for most of current AGR use cases.

Effective consent is another critical challenge when deploying AGR systems. Consent can only be effective if participants can understand the technology they will be subjected to and its limits. In our case, one of the author was always there to answer participants questions regarding the system but this cannot be the case if such a system were to be deployed.
Even with our in-depth explanation on AGR, such as how the algorithm works conceptually, and what data was used for training the model, some participants, both technical and non-technical, asked for more implementation details, such as more precise demographic details (i.e., race, gender, and age), size of the training data set and accuracy of the AGR algorithm on cis-gender people. This highlights the importance of providing context for users of AGR systems. 
However, potential users can come from diverse technical and cultural backgrounds and could have drastically different levels of understanding and diverging interpretation in terms of the bias and implications ingrained within AGR and its uses. The responsibility falls upon the application creators to provide  appropriate factual information in an easy to understand manner together with explanation for all the relevant flaws and bias of the system. We believe that this should be created on a per application basis by carefully examining (non-)target audiences and evaluating how the application might or not impact their perception. 

\subsubsection{Potential physical and mental health implications}
In the interviews, seven participants either expressed an interest or directly requested an access to the passing feedback. Some participants were interested in understanding their transition progress, others wanted to see it out of curiosity. However, a few participants also pointed out the dangerousness of displaying this specific metric.

These observations pose multiple potential mental health implications that should be acknowledged and taken into account. It is unknown how the participants would feel if they could see the passing feedback being less than favorable. And while we did inform the participants of the inherent technical challenge and inaccuracy that comes with repurposing AGR as a feedback mechanism for passing, some participants still felt that it performs better than human beings and felt obligated to trust it. It is genuinely possible that participants feel that their efforts to pass are invalidated by the system.

Secondly, how the system might impact the user's makeup decision remains unknown. As P9 suggested, it is possible that the system suggestions might oppress or discourage makeup options with worse feedback. It may also pushes users to strictly follow the recommendation, which we witnessed when P13 expressed her desire to depend on the system entirely for her makeup selection despite the feedback not aligning with her expectation. This can be especially more true for people who have a strong desire to pass. We think that long-term studies on potential psychological impact such a system can have on participants should be carried out before publicly releasing it.

\subsection{Design implication}
We designed our tool in a gender-neutral manner without targeting any specific gender identity. The makeup templates were heavily inspired by SNOW, a commercial application that does not strictly focus on makeup for women, and we did our best to select templates that work for Asian men (e.g. cheek contouring) and women (e.g. chin contouring) respectively, by looking at trans focused makeup websites. However, by not actively implementing trans-men oriented features, we admittedly missed out a valuable opportunity to address the specific needs of trans-men, such as virtual mustache and hair style. Future application developers should not only examine gender neutral features but also those that are specific to certain gender identities.

In addition, our approach to use makeup to assist trans-men passing faces several challenges that have to be overcome if this type of tool is to gain wild adoption in the community. While some of the trans-men participants are simply not interested in makeup, P9 and P10 both expressed the desire to wear makeup, but cited the strong association between makeup and womanhood, as well as the lack of male public figures wearing makeup as a strong discouragement for them to use makeup. Future developers of AGR systems should carefully examine not only the need of their potential audience, but also of non-audiences as other specific needs are surely yet to be met. 

%% file: 9-conclusion.tex
\section{Conclusion}
\label{sec:conclusion}

In this paper, we explored whether and how we can use machine learning and AGR systems to assist the Japanese transgender community explore makeup for passing. We first conducted a digital ethnography to better understand the specific needs of the transgender community in Japan. Based on the findings from the ethnography, we designed and developed a virtual makeup prototype to assist transgender individuals with passing using an AGR system. We then used this system with 15 participants living in Japan to inquire into their perception of an AGR-enabled virtual makeup system. We first analyzed their experiences with existing virtual makeup and AGR systems and found that some of them already appropriate existing AGR systems for their own needs. We also found that trans-specific needs for virtual makeup systems are unmet with existing technologies that focus mainly on cis women. In the second part of the study, we explored how participants perceived our virtual makeup system and found that while the system has technical limitation, participants most interested in passing found that the makeup feedback could be a tool that assists them with choosing and improving makeup for passing. This study leads us to conclude that under strict conditions and in specific contexts (i.e., Japanese), the AGR-based systems can potentially be appropriated by and benefit transgender individuals. We argue that the issues with AGR lay less in the technology itself than in how it is conceptualized, developed and deployed.

%% file: 10-acknowledgement.tex
\section{Acknowledgement}
We warmly thank all the participants for their time and comments regarding our system. We also thank the anonymous reviewers who helped us improve the paper. 
This work was supported by JST CREST Grant Number JPMJCR17A1, Japan.